\documentclass[12pt]{article}

\usepackage{amsmath}
\usepackage{amsfonts}

\usepackage{graphicx}

\usepackage[letterpaper,margin=1in]{geometry}

\renewenvironment{abstract}
	{\quotation}
	{\endquotation}

\date{}

\makeatletter
\renewcommand{\fnum@figure}{\textbf{Figure \thefigure}}
\renewcommand{\fnum@table}{\textbf{Table \thetable}}
\makeatother

\usepackage[numbers,sort&compress]{natbib}

\usepackage{url}

\usepackage{xcolor}
\newcommand{\mgp}[1]{\textcolor{black}{#1}}

\def\scititle{
	Near-Hamiltonian dynamics and energy-like quantities of next-generation neural mass models
}
\title{\bfseries \boldmath \scititle}

\author{
	Daniele~Andrean$^{1,2}$,
	Morten~Gram~Pedersen$^{1,3\ast}$ \and
	\small$^{1}$Department of Information Engineering, University of Padova, 35131 Padova, Italy.\and
	\small$^{2}$Department of Physics and Astronomy, University of Padova, 35131 Padova, Italy.\and
    \small$^{3}$Padova Neuroscience Center, University of Padova, 35129 Padova, Italy.\and
	\small$^\ast$Corresponding author. Email: mortengram.pedersen@unipd.it\and
}

\begin{document} 

\maketitle

\begin{abstract} \bfseries \boldmath
Neural mass models describe the mean-field dynamics of populations of neurons. 
In this work we illustrate how fundamental ideas of physics, such as energy and conserved quantities,
can be explored for such models.
We show 
that time-rescaling renders
recent next-generation neural mass models Hamiltonian in the limit of a homogeneous population or strong coupling.
The corresponding energy-like quantity provides
considerable insight into the model dynamics
even in the case of heterogeneity, 
\mgp{and explain for example why orbits are near-ellipsoidal and predict spike amplitude during bursting dynamics}.  
%
We illustrate how these energy considerations provide a possible link between neuronal population behavior and energy landscape theory, which has been used to analyze data from brain recordings.
Our introduction of near-Hamiltonian descriptions of neuronal activity 
could permit the application of 
highly developed 
physics 
theory
to get insight into brain behavior. 
\end{abstract}

\noindent

Whereas physics is based on
fundamental laws such as energy conservation, similar general principles describing the dynamics of physiological and neurological systems are still scarce.
However, 
energy considerations have also entered areas of biology and are for example widely used for the study of protein folding \cite{onuchic97,roney22}.
In neuroscience, Hopfield \cite{hopfield82} used  energy functions 
constructed \textit{ad hoc}
 to study the stability of equilibrium states in his seminal work on memory forming networks.
Further, energy landscapes -- a concept borrowed from statistical physics -- are used to analyze, e.g., EEG \cite{klepl21,watanabe21} and fMRI \cite{watanabe14} brain recordings, and have provided insights into the effects of transcranial magnetic stimulation \cite{watanabe21} and disease \cite{krzeminski20,klepl21,masuda25}. We stress that the energies considered in these setting bare no relation to, e.g., metabolic energy, but are mathematical structures based on ideas from physics.
Importantly, brain energy landscapes are derived directly from data as a convenient way to synthesize and reason about the experimental recordings \cite{masuda25}, but have no clear connection to the underlying neuronal populations. 


Mean-field neural models aim to describe the overall behavior of a population of neurons \cite{bick20}. 
Recently, so-called
next-generation neural mass models were introduced and shown to accurately describe the mean behavior of a population of quadratic integrate-and-fire (QIF) neurons coupled synaptically \cite{montbrio15,coombes23}.
These models can produce complex behavior including population oscillations, 
and even bursting in the presence of 
synaptic depression
or spike-frequency adaptation \cite{gast20,ferrara23}.
Mathematical analysis of these models is challenging, and it is not clear how to link their dynamics to the energy considerations mentioned above.
In particular, whether these models can be approximated by Hamiltonian descriptions and consequently possess near-conserved 
energy-like
quantities, which could provide a link to energy-landscape theory, has not been investigated.



\section*{Near-Hamiltonian description of a next-generation neural mass model with fast synapses}

In the case of global, instantaneous coupling between a population of QIF neurons, and assuming that the individual excitabilities
follow a Lorentzian distribution,
the resulting next-generation neural mass model can be written as \cite{montbrio15}
\begin{equation}\label{2dmodel}
    \dot r = \frac{\Delta}{\pi} + 2rv\ , \quad \dot v = v^2 -\pi^2r^2 \pm Jr+\eta + I(t),
\end{equation}
where overdots indicate differentiation with respect to time $t$, $r\geq 0$ is the average firing rate, $v$ models the population voltage,
$J>0$ is the synaptic strength,
and $I(t)$ is a time-varying applied current.
%
The Lorentzian distribution is characterized by its median $\eta$ and the 
half-width at half-maximum 
$\Delta$, which is thus
a measure of population heterogeneity.
$\Delta$ may also account for noise in the individual neurons \cite{clusella24}. Depending on the sign of the term $\pm Jr$, the synapses may be excitatory ($+$) or inhibitory ($-$).
This neural mass model is asymptotically exact for a Lorentzian distribution and provides an excellent description of the population dynamics for other types of distributions \cite{montbrio15}.

We observe that rescaling by $R=r/J$, $V=v/J$, $T=Jt$ effectively reduces the number of parameters by one, as we obtain
\begin{equation}\label{2DMODEL}
     R' = \delta + 2 RV\ , \quad V' = V^2 -\pi^2R^2 \pm R+\lambda + i(T),
\end{equation}
where differentiation is with respect to $T$, $\delta = 
{\Delta}/({\pi J^2})$, $\lambda = \eta/J^2$ and $i(T)=I(T/J)/J^2$. 
In many works with interesting dynamics, $\Delta\ll J^2$, and hence $\delta \ll 1$. For example, 
bistability appears only for $\Delta/J^2<1/64$ (i.e., $\delta<0.005$), and 
it occurs over a moderately wide range of $\eta$ values only for $\delta<0.003$ \cite{montbrio15}.

Setting 
$i(T)=i_0$ in \eqref{2DMODEL} gives a system of the more general form
\begin{equation}\label{2DMODEL_K}
     R' =  \delta + 2 RV\ , \quad V' = V^2 - \pi^2R^2+ \Lambda(R),
\end{equation}
for which 
\begin{equation}\label{H_K}
    H(V,R) = \frac{V^2}{R}+\pi^2 R - \int^R \frac{\Lambda(r)}{r^2} dr
\end{equation} is conserved if $\delta=0$ and the integral is well defined. 

Rescaling time state-dependently to 
$\mgp{\theta}=T\cdot R^2$ in \eqref{2DMODEL_K} 
yields 
\begin{equation}\label{2DMODEL_K_ham}
     \frac{dR}{d\mgp{\theta}
     } =  \frac{\delta}{R^2} +  \frac{2V}{R}\ , \quad \frac{dV}{d\mgp{\theta}
     } = \frac{V^2}{R^2} - \pi^2+ \frac{\Lambda(R)}{R^2},
\end{equation}
which is a perturbed Hamiltonian system with $H$ being the Hamiltonian.
Indeed, for $\delta=0$, $\frac{dR}{d\mgp{\theta}
     } = \frac{\partial H}{\partial V},\ \frac{dV}{d\mgp{\theta}
     }= - \frac{\partial H}{\partial R}$. Thus, $V$ is the generalized momentum and $R$ the generalized position. 
Moreover, 
$U(R):=\pi^2 R - \int^R \frac{\Lambda(r)}{r^2} dr$ 
can be seen as the potential energy of the system, and 
$K:=\frac{V^2}R = \frac14 R\cdot \big(\frac{dR}{d\mgp{\theta}
     }\big)^2$
as the kinetic energy, where the ``mass" $m(R)=R/2$ is
position-dependent \cite{cruz13}. 
We note that the perturbation term $\delta/R^2$ in \eqref{2DMODEL_K_ham} may become large if $R$ is of order $\sqrt{\delta}$.

Setting $H(V,R)$ equal to a constant $H_0$ yields closed curves in the $(V,R)$ phase space that are symmetric with respect to the $R$ axis and ``circular" in the sense that for a given value of $R$ correspond exactly two values of $V$ with opposite sign, whenever we can solve $H(V,R)=H_0$ for $V$. For $\delta=0$, these closed curves are the trajectories of the system. 
Moreover, since $\frac{dH}{dT}= -\frac{\delta}{R^2}\frac{dV}{dT}$, we see that, for $0<\delta\ll 1$, $H$ increases slightly along the top of these curves where $V$ decreases.
Along the lower part of the curves, where $V$ increases, 
the decrease in $H$ is greater than this small increase,
since here $R$ is smaller. 
Consequently, overall the systems spirals  down along the $H$ energy surface towards the equilibrium.

In the case of \eqref{2DMODEL}, i.e., for $\Lambda(R)=\pm R+\lambda+i_0$, we obtain 
\begin{equation}\label{H_K2}
H(V,R)=\frac{V^2}{R}+\pi^2 R \mp \log(R)+ \frac{\lambda+i_0}R.
\end{equation}
For $\delta=0$, 
the trajectories of the system are the closed curves 
$$
V=\pm \sqrt{R H_0-\pi^2 R^2 \pm R\log(R)- \lambda-i_0},
$$
reflecting the behavior seen in simulations of 1000 identical, coupled QIF neurons \cite{laing18}.
We expect the model with $0<\delta\ll 1$ to follow these curves approximately as it spirals 
down along the $H$ energy surface
towards the stable node, which is confirmed by numerical simulations (Fig.~\ref{fig:RV_Montbrio}). 
Thus, the behavior of the neuronal population can be likened to a ball that moves in a 1-dimensional potential $U$ in the presence of friction, due to $0<\delta\ll1$, which causes the system to perform damped oscillations towards the minimum of $U$.

\section*{Energy considerations explain current-induced bursting}

The model 
can produce bursting when the applied current is slowly varying \cite{montbrio15}, here 
$i(T) = A \sin(2\pi \omega T)$ (Fig.~\ref{fig:RVI_Montbrio}A).
The energy considerations introduced above help us to understand the behavior during the active phase.
We consider an excitatory population. 
The current causes a slow modulation of the potential energy surface $U=\pi^2 R-\log R + \frac{\lambda +i(T)}{R}$, which for 
$i(T)<-\lambda -\frac1{4\pi^2}$ becomes strictly increasing and no longer possesses a local minimum (Fig.~\ref{fig:RVI_Montbrio}B). 
At this point, the system ``falls off" the potential surface and $R$ tends to zero, which ends the active phase. 
Since $\frac{dU}{dT}=U'(R)\frac{dR}{dT}$,
the drop in $U$ stops  -- for non-zero $\delta$ -- before $R=0$ when $R'=0$.

The fast oscillations occurring during the active phase can be estimated from the 
closed, energy-constant ($H(V,R)=H_0$) curves introduced above (Fig.~\ref{fig:RVI_Montbrio}C).
Since these curves are characterized by the couple $(H_0,i_0)$,  they can be seen as a critical manifold (i.e., for $\delta=0$) over the $(H,i)$ plane, which we thus expect to provide a good description of the fast dynamics as $(H,i)$ change slowly for $0<\delta\ll 1$ and $i$ non-constant.
For example, the maximum and minimum values of the fast spikes in $R$ can be found by solving $R \cdot H(T)=\pi^2 R^2 - R\log(R) +\lambda+i(T)$ for $R$, 
corresponding to the upper and lower points of the 
energy-constant
curves where the kinetic energy $K$ is zero, which provide good estimates of the amplitude of the simulated fast oscillations (Fig.~\ref{fig:RVI_Montbrio}D).
Here, $H(T)=H\big(V(T),R(T)\big)$ is the value of $H$ calculated at time $T$. 

During rest with low firing rate ($R\approx 0$) and low mean voltage ($V<0$), the ``energy" of the system $H$ increases quickly (``regenerates"; Fig.~\ref{fig:RVI_Montbrio}A), preparing for a new round of activity. 
As $R\to 0$ when the active phase ends, the system crosses and stays below the $V$ nullcline, causing $V$ to decrease and $H$ to increase, because of the relation $\frac{dH}{dT}=-\frac{\delta}{R^2}\frac{dV}{dT}$. In energy terms, the rapid drop in potential energy is compensated by an increase in kinetic energy $K$, and in addition $\delta>0$ here acts a source of energy, rather than as energy-consuming friction.
When $i$ increases again, eventually reintroducing the local minimum separated from $R\approx 0$ by a potential barrier, the system has enough kinetic energy to overcome this barrier and enter the basin where damped oscillations occur (Fig.~\ref{fig:RVI_Montbrio}B).

The system thus possesses two energy-states in its ``energy landscape". One has high potential energy $U$, whereas the other is at low potential energy $U$ but high kinetic energy $K$. These states coexist, separated by an energy barrier, for a range of $i$ values and correspond to the two stable equilibria with $V\approx 0$ and $R\approx 0$, respectively \cite{montbrio15}. 
As in energy landscape theory, an external stimulus, or noise, may push the system from one energy state to the other, and, moreover, a slowly varying input can modulate the energy landscape dynamically.

\section*{Bursting in an inhibitory population}
\begin{figure}[tb]
    \centering
    \includegraphics[width=0.8\linewidth, trim= 0 25cm 0 0, clip]{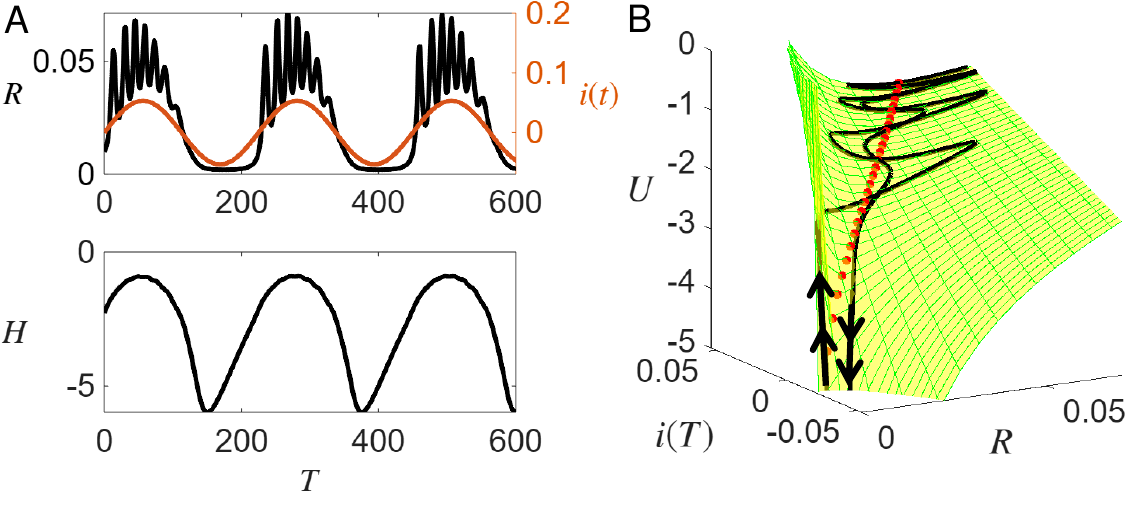}
    \caption{\textbf{Bursting in an inhibitory population 
    in response to a slowly varying current.}
    A) Firing rate $R$ (upper) in an inhibitory population stimulated by a sinusoidal current 
    $i(T) = A \sin(2\pi \omega T)$ (lower).
    B) During the active phase of bursting, the system oscillates around the local minimum of the potential energy $U$, both for increasing and decreasing $i$.
    The local minimum disappears at $R=0$ for $i=-\lambda$, and the system then goes towards the low-potential equilibrium with $R\approx 0$. 
    Red dots show local minima of $U$. 
    Arrows indicate the direction of movement along the trajectory.
    Simulation parameters are $\delta = 0.0007$, $\lambda = 0.0222$, $A=0.0533$, $\omega=0.0044$.}
    
    \label{fig:RVI_inhib_Montbrio}
\end{figure}

In the case of inhibitory coupling, the system \eqref{2DMODEL} has a single equilibrium with $R>0$, which is always stable \cite{devalle17}. This fact corresponds to the potential energy surface, for $R>0$, being either U-shaped with a single minimum (for $\lambda+i >0$), or being strictly increasing (for $\lambda+i\leq 0)$. 
In the case of a single minimum, the system typically oscillates around the energy state (corresponding to the equilibrium being a stable focus), whereas for $\lambda+i\leq 0$ the energy state at  $R\approx 0$ has low potential energy, similarly to the behavior seen for the excitatory population (Fig.~\ref{fig:RVI_Montbrio}). 
These observations suggest, as confirmed by simulations (Fig.~\ref{fig:RVI_inhib_Montbrio}), that a slowly varying current may also induce bursting in the inhibitory population by dynamically changing the energy landscape, causing the system to alternate between the high and low potential energy states, which are however never coexisting.

\section*{
Quasi-Hamiltonian analysis of bursting with spike-frequency adaptation}
\begin{figure}[htb]
    \centering
    \includegraphics[width=0.7\linewidth]{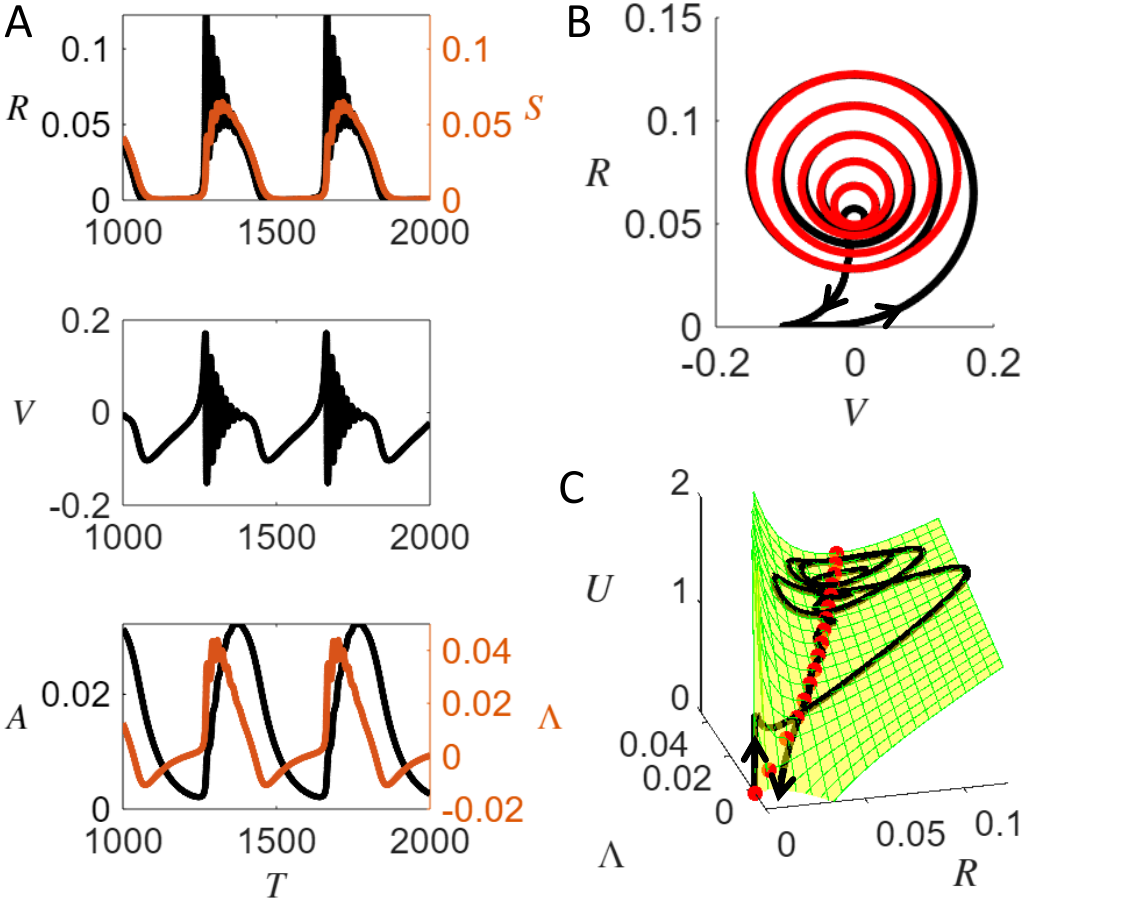}
    \caption{\textbf{Energy considerations for bursting in the SFA model.}
    (A) Time series of model variables. 
    (B) 
    The trajectory (black) follows constant-energy ellipses (red) 
    in the $(R,V)$ phase-plane during the active phase of bursting.
    (C) \mgp{The} potential energy surface \mgp{is geometrically similar to the one of the model \eqref{2DMODEL} with fast inhibitory synapses  
    (Fig.~\ref{fig:RVI_inhib_Montbrio}B), and} the system follows the valley of  minima 
    during the active phase of bursting.
    Simulation parameters are $\delta = 6.4961 \cdot 10^{-5}$, $\lambda = 0.002$, $\epsilon=0.0143$, $\rho=0.0714$, $\beta=0.7143$. \\
    }
    \label{fig:Energy_surf_Olmi}
\end{figure}

We now study the model by Ferrara et al. 
\cite{ferrara23}, 
which extends the model \eqref{2dmodel} by including exponential synaptic coupling \cite{devalle17}, modeled by the variable $s$, and spike-frequency adaptation (SFA) through the variable $A$ (see also \cite{gast20}). 
Rewriting this model using the transformations above and $S=s/J$, $a=A/J^2$, 
we obtain
\begin{eqnarray}
    R' &=& 
    \delta +2RV, \label{dr} \\
    V' &=& V^2 - (\pi R)^2 +\lambda \pm S -a, \label{dv}\\
    S'&=& \rho(R-S), \label{ds}\\
    a'&=& \epsilon (\beta R-a) \label{dA},
\end{eqnarray}
where, compared to \cite{ferrara23}, $\beta=\alpha/J$,  $\epsilon=1/(J\tau_A$), and $\rho=1/(J\tau_s)$.
With appropriate parameters and excitatory synapses, the model can exhibit bursting behavior (Fig.~\ref{fig:Energy_surf_Olmi}), 
whereas inhibitory coupling typically produces spiking \cite{ferrara23}. 

To understand the dynamics during the active phase of bursting in the case of excitatory synapses, we set 
$\delta=0$ and treat 
$\Lambda:=\lambda+ S-A$ as a constant,
\mgp{since it changes little during the active phase (Fig.~\ref{fig:Energy_surf_Olmi}A). 
From \eqref{H_K} we then obtain that},
\begin{equation}
    \label{H}
    H=\frac{V^2}{R} + \pi^2 R +\frac{\Lambda}{R} 
\end{equation}
is a conserved quantity for \eqref{dr}-\eqref{dv}.
This means that, for $\delta=0$ and $\Lambda$ constant, orbits are ellipses described by
\begin{equation}\label{ellipse}
    R_0^2= V^2 +\pi^2(R-r_0)^2, \qquad r_0 = \frac{H}{2\pi^2},\quad R_0 = \sqrt{\frac{H^2}{4\pi^2}-\Lambda} = \sqrt{(\pi r_0)^2 -\Lambda}.
\end{equation}

\begin{figure}[htb]
    \centering
    \includegraphics[width=0.7\linewidth]{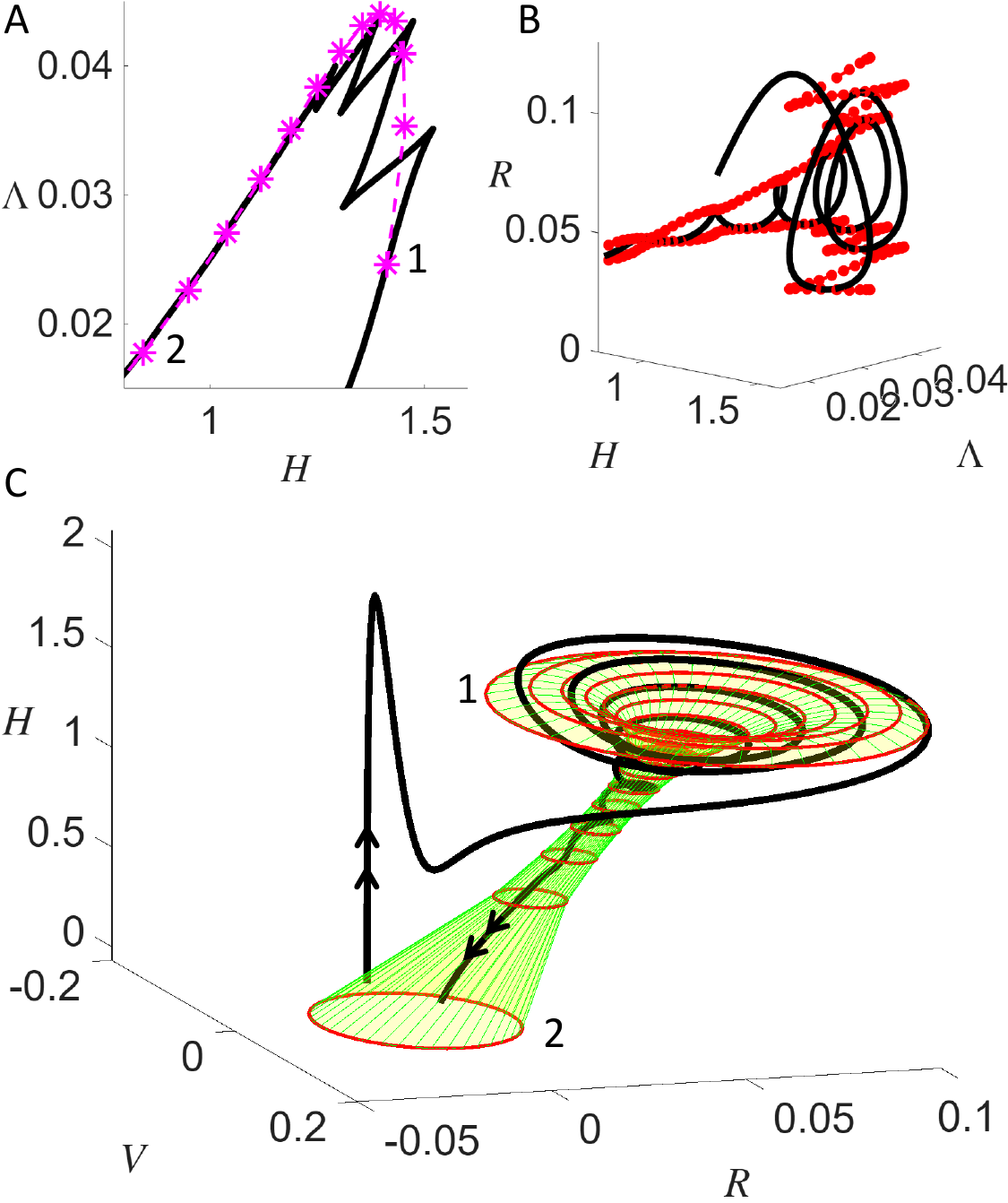}
    \caption{\textbf{Minima/maxima and averaging explain the dynamics of the SFA model.}
    A) trajectory of the model projected onto the $(\Lambda,H)$ phase-plane (black). The \mgp{purple asterisks} represent 
    the 
    solution of the 
    \mgp{discrete, averaged system \eqref{dK5}-\eqref{dH5}} starting from point 1 and ending in point 2. 
    B) Trajectory of the \mgp{SFA model \eqref{dr2}-\eqref{dA2}} 
    in the $(R,\Lambda,H)$ phase-space (black) with the maxima and minima of the ellipses composing the critical manifold (red). The simulated trajectory follows this critical manifold closely.
    C) 
    The ellipses, \mgp{defined by the asterisks in panel (A) and \eqref{ellipse},}
    form a funnel (green), which explains the model trajectory (black) well. The system travels down the funnel following constant energy ellipses. 
    Arrows indicate the movement direction along the trajectory. 
    \mgp{The ellipses indicated by 1 and 2 correspond to the 
    labeled} points in the $(\Lambda,H)$ phase-space in panel (A).
        Simulation parameters are $\delta = 6.4961 \cdot 10^{-5}$, $\lambda = 0.002$, $\epsilon=0.0143$, $\rho=0.0714$, $\beta=0.7143$.}
    \label{fig:minmaxavg_Olmi}
\end{figure}

\mgp{Similarly to the case of current induced bursting studied above, the} 
assumption of $\Lambda$, and hence $H$, being constant, yields a critical manifold composed of ellipses described by \eqref{ellipse} over the $(\Lambda,H)$ plane. We constructed the part of the critical manifold lying over the orbit projected onto the $(\Lambda,H)$ plane.
The simulated trajectory follows this critical manifold closely (\mgp{Fig.~\ref{fig:minmaxavg_Olmi}}B), which explains why the orbit projected onto the $(V,R)$ plane shows 
\mgp{near-}ellipsoidal oscillations \mgp{(Fig.~\ref{fig:Energy_surf_Olmi}B)}. 



We would like to understand the dynamics of the system in the $(\Lambda,H)$ plane, and thus how the system moves along the critical manifold composed of ellipses. 
We first note that the original system \eqref{dr}-\eqref{dA} can be rewritten as 
\begin{eqnarray}
    R' &=& \delta +2RV, \label{dr2}\\
    V' &=& RH - 2\pi^2 R^2 \label{dv2},\\
    \Lambda' &=& \pm \rho (R-\Lambda+\lambda -a) - \epsilon (\beta R-a) \nonumber \\
                  &=& (\pm \rho -\epsilon\beta) R \pm \rho (\lambda - \Lambda) - (\pm \rho-\epsilon ) a \label{dK2}\\
    a' &=& \epsilon (\beta R- a) \label{dA2},
\end{eqnarray}
with $H$ given by \eqref{H}. 

For $\delta=0$ and after rescaling of time to $\tau= 2RT$, we see that \eqref{dr2}-\eqref{dv2} become
\[
\frac{d^2R}{d\tau^2} =\frac{dV}{d\tau} = \frac{H}2 - \pi^2 R,
\]
which has solution (assuming $H$ constant, $V(0)=0$, $\frac{dV}{d\tau}>0$, i.e., starting from the lower point of the ellipse described by \eqref{ellipse})
\begin{equation}
    \label{rtau}
    R(\tau) = \frac{H}{2\pi^2} - \frac{R_0}{\pi} \cos(\pi \tau) = r_0 - \frac{R_0}{\pi} \cos(\pi \tau) \ , \quad V(\tau)=R_0 \sin(\pi \tau).
\end{equation}
This \mgp{exact} expression for the dynamics along the ellipses allows us to make explicit \mgp{averaging-type} calculations for how $\Lambda$, $a$ and $H$ can be expected to change during half a turn along the ellipses, i.e., for $n\leq \tau < n+1$ with $n\in \mathbb N$. In this way we obtain \mgp{the following} 
discrete system in the $(\Lambda,a,H)$ subspace \mgp{(see Supplementary Material)},

\begin{eqnarray}
    \Lambda_{n+1}  &=&  \Lambda_n+\frac{\pm \rho -\epsilon\beta}2 + \frac{\pm\rho (\lambda-\Lambda_n) - (\pm\rho-\epsilon)a_n}{2}
    \frac{\pi}{\sqrt{\Lambda_n}}\  , \label{dK5}\\
    a_{n+1} &=& a_n+ \frac{\epsilon \beta}{2}  -\frac{\epsilon a_n}{2} \frac{\pi}{\sqrt{\Lambda_n}}  \label{dA5}\ ,\\
    H_{n+1}  &=& H_n+\frac{(\pm\rho -\epsilon\beta)}2 \frac{\pi}{\sqrt{\Lambda_n}}  + 
    \frac{\pm\rho (\lambda-\Lambda_n) - (\pm\rho-\epsilon)a_n}{4} \cdot
    \frac{\pi{H_n}}{\sqrt{\Lambda_n^3}}\ . \label{dH5}
\end{eqnarray}
The system \eqref{dK5}-\eqref{dH5} is thus a closed system, predicting how the complete system goes from one (semi-)ellipse to the next along the critical manifold (Fig.~\ref{fig:minmaxavg_Olmi}A),
\mgp{which provides an excellent description of the overall bursting dynamics (Fig.~\ref{fig:minmaxavg_Olmi}C).
For example, the fact that initially $H$ is almost constant while $\Lambda$ increases (Fig.~\ref{fig:minmaxavg_Olmi}A), together with \eqref{ellipse}, explains why the centers of the ellipses move very little during the beginning of the active phase whereas the radii decrease.}

\section*{Energy analysis of bistability and switching in two inhibitory populations}
To illustrate how the energy-considerations can be applied to more than one population,
we modeled mutual inhibition between two neuronal populations, following \cite{montbrio15}, as
    $R_k' =  \delta + 2 R_kV_k$, $V_k' = V_k^2 - \pi^2R_k^2  - R_k+\lambda - J_{c}R_l/J + i_k$ , $k,l=1,2$, $k\neq l$, where $i_k$ is a current stimulus given to population $k$.
Mutual inhibition 
naturally leads to bistability where one 
population is active and firing ($R>0$) whereas the other one is inhibited and virtually silent ($R\approx 0$). 
The total potential energy of the system, i.e., the sum of potential energies $U_1$ and $U_2$ in the two populations, defines the energy landscape of the system, which has two valleys corresponding to the drop in $U_k$ as $R_k\to 0$  (Fig.~\ref{fig:switching}, green surface; compare with Fig.~\ref{fig:RVI_inhib_Montbrio} for negative input).
These two symmetric energy states correspond to the two stable equilibria, where one (non-inhibited) population is in the high-potential energy state where it inhibits the other population. The resulting inhibitory (negative) current makes the inhibited population go to the low-potential energy state at $R\approx 0$.

We explored whether an external stimulus could 
induce a transition of the system from one energy state to another, 
similarly to how transcranial magnetic stimulation has been 
found to influence state transitions in the context of energy landscape theory \cite{watanabe21}.
Adding a brief current to the inhibited population (
population 1) sitting at the low-energy state
caused the total-potential-energy 
surface to move upwards, and in addition increased the kinetic energy of population~1 
(by increasing $V_1$), which allowed the system to move uphill along the energy surface, increasing $R_1$ (Fig.~\ref{fig:switching}, black curve).

If the current was sufficiently strong, the systems switched as $R_1$ started to inhibit the second population. Thus, if the system had passed the ridge at $R_1=R_2$ on the energy surface when the current was removed, the system would slide along the energy surface towards the energy state where population 1 inhibits population 2 ($R_1>0, R_2\approx 0)$, i.e., the current input caused the populations to switch energy states
(Fig.~\ref{fig:switching}, black curve). 
If, on the contrary, the current was insufficient to cross the energy barrier along the ridge of the energy surface, the system would return to the original state with low potential energy for population 1 ($R_1\approx 0, R_2>0$) (Fig.~\ref{fig:switching}, red curve).
Interestingly, 
the switch in energy states could occur 
even if the trajectory had not 
passed the ridge in Fig.~\ref{fig:switching} at the end of the stimulus, i.e.,
if $R_2>R_1$, when the current was removed (Fig.~\ref{fig:switching}, blue curve).
In this case, the kinetic energy played a role in the switch akin to a ball that rolls uphill and is able to cross a modest energy barrier.

\section*{\mgp{Conclusions}}\label{sec12}

\begin{figure}
    \centering
    \includegraphics[width=0.8\linewidth, trim=0cm 26cm 0cm 0cm, clip]{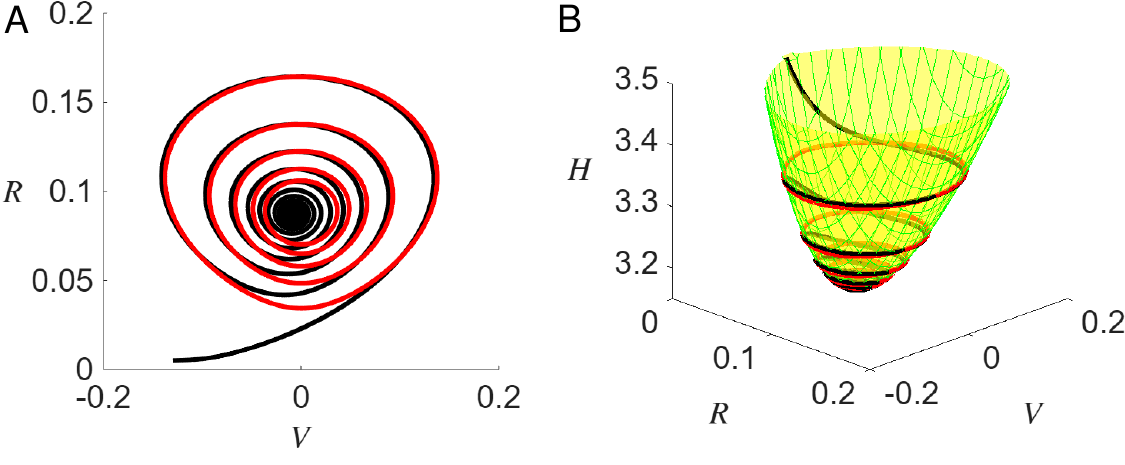}
    \caption{\textbf{System dynamics is well described by energy-constant curves.} 
    A) The  trajectory (black) of system \eqref{2DMODEL} with excitatory coupling 
    closely follows the energy-constant curves (red) as it spirals towards the equilibrium.
    B) The trajectory (black) spirals down the energy surface (yellow) towards the equilibrium. Red curves in panel A correspond to the horizontal intersections (red) with the energy surface. 
    Parameters are $\delta=0.0014$, $\lambda=-0.0222$, and $i(T)=0.2$.
    }
    \label{fig:RV_Montbrio}
\end{figure}

\begin{figure}[tb]
    \centering
    \includegraphics[width=0.8\linewidth, trim= 0 19cm 0 0, clip]{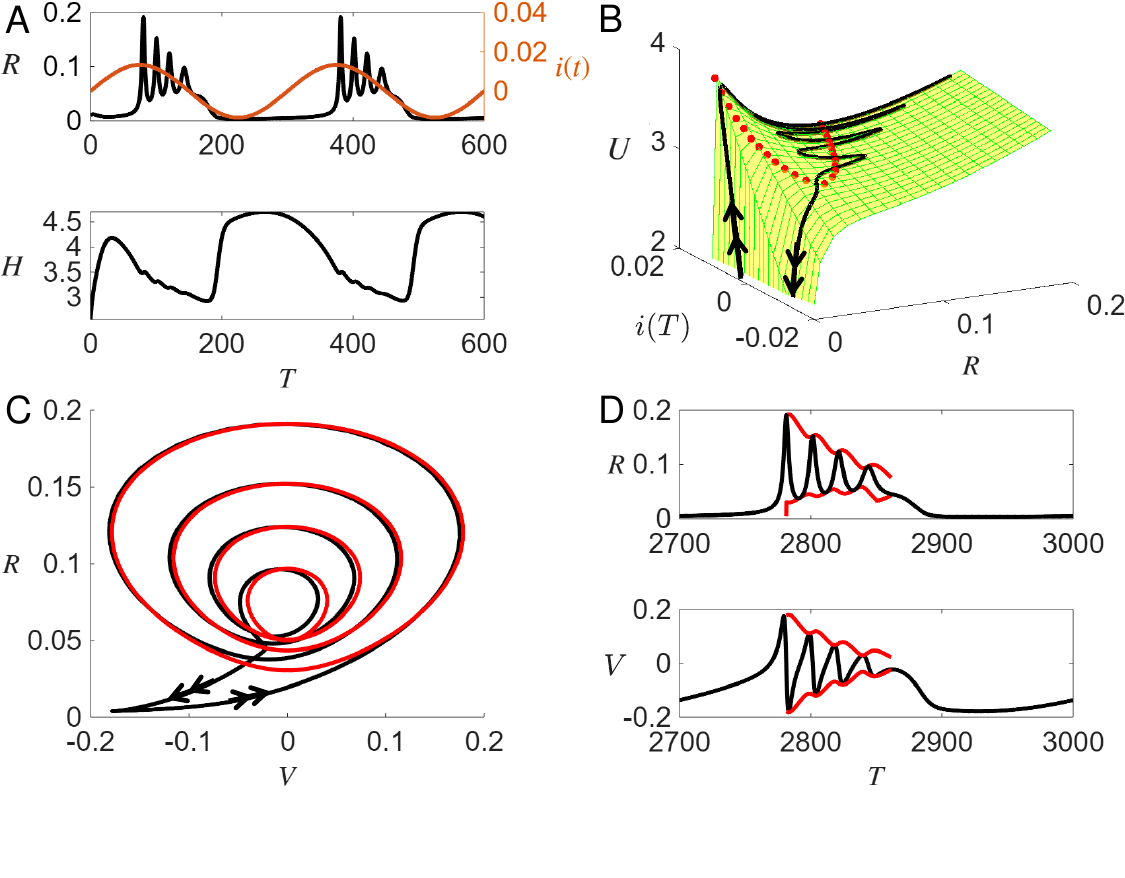}
    \caption{\textbf{Bursting population behavior in response to a slowly varying current.}
    A) Firing rate $R$ (upper) in an excitatory population stimulated by a sinusoidal current 
    $i(T) = A \sin(2\pi \omega T)$ (lower).
    B) During the active phase of bursting, the system oscillates around the local minimum of the potential energy $U$ as $i$ decreases.
    The local minimum disappears when it meets the local maximum in a cusp catastrophe for $i=-\lambda-1/(4\pi^2)\approx -0.0031$, and the system then goes towards the low-potential equilibrium with $R\approx 0$. 
    Red dots show local minima/maxima (for fixed $i$). Arrows represent the direction of movement along the trajectory.
    C) Energy-constant curves well approximate the system trajectory in the $(R,V)$ phase-plane. Arrows represent the direction of movement along the trajectory.
    D) The amplitude of small-amplitude oscillations is well predicted by energy considerations. 
    In all simulations $\delta = 0.0014$, $\lambda = -0.0222$, $A=0.0133$, $\omega=0.0033$).
    }
    \label{fig:RVI_Montbrio}
\end{figure}

Our finding that next-generation neural mass models are near-Hamiltonian allowed us to introduce 
the energy-like quantity $H$, which is conserved in the limit $\delta\to 0$. 
We obtained an explicit expression of how $H$
depends on the mean firing rate $R$ and the mean voltage $V$ in the population, and showed that the energy-constant curves provide excellent approximations of the behavior of the system as it spirals towards the stable focus (Figs.~\ref{fig:RV_Montbrio} and \ref{fig:RVI_Montbrio}). 

Further, the Hamiltonian viewpoint leads naturally to concepts such as potential and kinetic energy for the system. 
We found that 
the neural mass system should be 
considered as having space-dependent mass \cite{cruz13} increasing linearly with $R$. 
The potential energy $U$ of the neural mass model is a function of the firing rate only, similar to how the potential energy of a mechanical system is a function of position. 
This correspondence suggests that 
equilibrium points can be found as the minima of the potential-energy graph (or surface), see Fig.~\ref{fig:RVI_Montbrio}.
However, in contrast to mechanical systems, the neural mass system may move uphill so that stable equilibria appear on the down-slope of the graph, in particular for $R\approx 0$ where the non-Hamiltonian perturbation becomes large. 

\begin{figure}[tb]
    \centering
    \includegraphics[width=0.6\linewidth, trim= 0cm 6cm 20cm 0cm, clip]{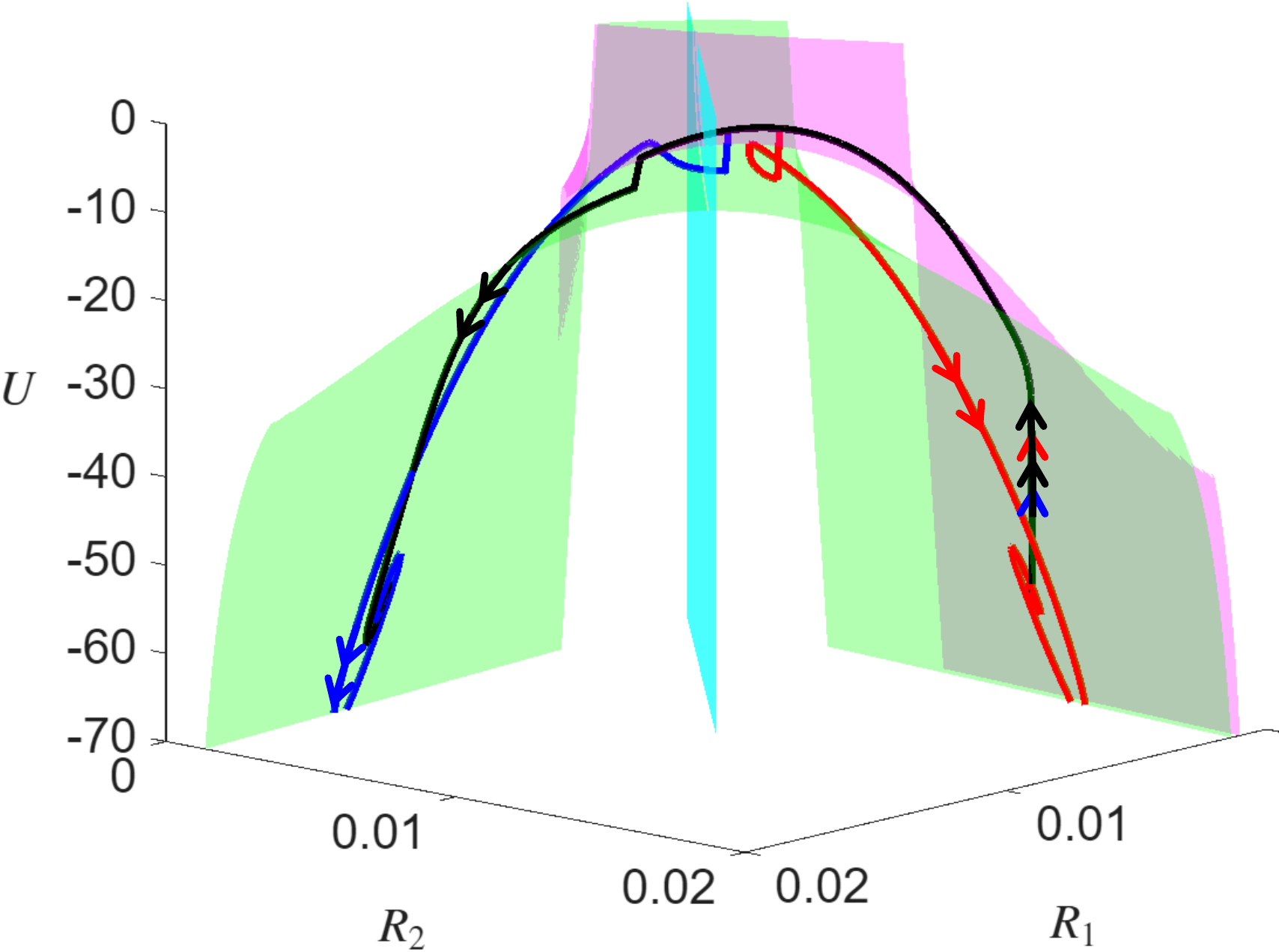}
    \caption{\textbf{A current stimulus can induce switching behavior in two coupled populations.}
    The total potential energy surface of two mutually inhibitory populations is shown in green.
    Initially, the system is sitting at the energy state at $R_1\approx 0$. 
    A current stimulus, of amplitude $i=0.0201$ and duration $100$ time units, is given to the inhibited population ($R_1$), which raises the energy surface (magenta) and causes an increase in kinetic energy, allowing the system to move up along the energy surface (black trajectory). When the stimulus ends, the system returns to the no-stimulus (green) energy surface, but now on the other side of the surface where $R_1>R_2$ (the cyan vertical plane indicates $R_1=R_2$). Here, it moves down along the surface towards the energy state at $R_2\approx 0$.
    In case of a slightly shorter stimulus (duration $94$ time units, red), the system has not passed the ridge when the stimulus ends and the trajectory jumps down to the no-stimulus (green) energy surface. Consequently, the system returns to the initial energy-state at $R_1\approx 0$. Stimuli with intermediate duration (e.g., $98$ time units, blue) may cause the system trajectory to jump down before the $R_1=R_2$ plane. However, due to remaining inertia, the system is able to cross the ridge and go down to the other state. Note that the initial trajectory is the same for all 3 simulations as only the duration of the stimulus is varied. 
    Similar effects can be obtained by varying the  amplitude of the stimulus, but the with-stimulus energy surface would change according to the amplitude.
    Simulation parameters are $\delta = 0.0004$, $\lambda = 0.0187$, $J=20$, $J_c=100$.}
    \label{fig:switching}
\end{figure}

If co-existing, these equilibrium states are separated by an energy barrier, very similar to how energy-states in neuronal energy landscape theory \cite{masuda25} are assumed to be separated by energy barriers. Whereas energy landscape theory estimates the energy barriers and states by analyzing how data from brain recordings transition between different meta-stable configurations, we derived the energy landscape directly as the potential-energy surface of the neural mass model. We illustrated how this picture can be used to investigate transitions between energy-states in response to a stimulus (Fig.~\ref{fig:switching}), which can be seen as a highly simplified version of the modifications of the energy landscape in response to transcranial magnetic stimulation \cite{watanabe21}.

\mgp{For the model with spike-frequency adaptation, we explained why orbits are (near) ellipsoidal  (reminiscent of Kepler's first law for planetary orbits!) during the active phase of bursting. Using time rescaling, we found an explicit expression for the movement along these orbits, valid for $\delta=0$, which allowed us to perform averaging to predict the dynamics along the critical manifold and thus, e.g., the spike-amplitude during bursting.
The assumption of $\Lambda$ being near-constant was based on simulations and the fact that both $s$ and $A$ increase during the active phase, and hence their difference does not change much. It would be interesting to validate theoretically that $\Lambda$ is indeed a slow variable.
}

The Hamiltonian descriptions become more accurate as $\delta = \Delta/(\pi J^2)$ gets smaller.
This may happen either because the neuronal population is very homogeneous ($\Delta\approx 0$) or because the coupling strength $J$ is very large. In other words, a sufficiently large coupling may overcome heterogeneity in the population and make it 
a near-Hamiltonian system by causing the neurons to behave as if they were identical. 

Our results 
should be generalizable beyond
populations of QIF neurons with Lorentzian-distributed excitability parameters, since QIF neurons represent the normal form of the large class of type I conductance-based neuron models \cite{izhikevich07} and, moreover, the mean-field approximation is valid also for non-Lorentzian distributions \cite{montbrio15}.
Further, extending the ideas presented here to populations with, e.g., electrical synapses \cite{pietras19} or delayed coupling \cite{devalle18}
should be feasible and could provide insight into more realistic mean-field models of neuronal activity and, ultimately, brain behavior.




\clearpage

\clearpage 

%
\bibliography{QIF}
\bibliographystyle{sciencemag}

%
%
%
%
%
%


\newpage
\section*{Acknowledgments}
\paragraph*{Funding:}
No specific funding was received for this study.
\paragraph*{Author contributions:}
D.A. contributed to theory, wrote computer code, performed simulations, prepared figures, and revised the manuscript.
M.G.P. conceptualized research, developed theory, and wrote the paper.
\paragraph*{Competing interests:}
There are no competing interests to declare.
\paragraph*{Data and materials availability:}
No data was produced for this study. 
All figures can be reproduced with standard software solvers of ordinary differential equations.

\newpage

\renewcommand{\thefigure}{S\arabic{figure}}
\renewcommand{\thetable}{S\arabic{table}}
\renewcommand{\theequation}{S\arabic{equation}}
\renewcommand{\thepage}{S\arabic{page}}
\setcounter{figure}{0}
\setcounter{table}{0}
\setcounter{equation}{0}
\setcounter{page}{1} 


\begin{center}
\section*{Supplementary Material for\\ \scititle}

Daniele~Andrean,
	Morten~Gram~Pedersen$^{\ast}$ \and

	\small$^\ast$Corresponding author. Email: mortengram.pedersen@unipd.it\and

\end{center}

\subsection*{Averaging calculation for the model with spike-frequency adaptation}



For $\delta=0$, we rewrite \eqref{dK2}, $\eqref{dA2}$ and the expression for $H$ on the $\tau$ timescale to get 
\begin{eqnarray}  
    \frac{d\Lambda}{d\tau}  &=& \frac12(\pm\rho -\epsilon\beta) \pm \frac{\rho}{2R}(\lambda -\Lambda)-\frac{a}{2R}(\pm \rho-\epsilon ) \label{dK3}\\
    \frac{da}{d\tau} &=& \frac{\epsilon \beta}{2}  -\frac{\epsilon a}{2R} \label{dA3},\\
    \frac{dH}{d\tau}  &=& \frac1{2R}(\pm\rho -\epsilon\beta) \pm \frac{\rho}{2R^2}(\lambda -\Lambda)-\frac{a}{2R^2}(\pm \rho-\epsilon )   \label{dH3}
\end{eqnarray}
Assuming $\Lambda,a,H$ constant during a half-turn along the ellipses, we 
update these variables by calculating the average of $1/R$ and $1/R^2$ along the half-turn of the ellipse. That is, we define
\begin{eqnarray}
    \Lambda_{n+1}  &=& \Lambda_n+ \Big\langle\frac{d\Lambda}{d\tau}\Big\rangle_n 
    = \Lambda_n+\frac{\pm \rho -\epsilon\beta}2 + \frac{\pm\rho (\lambda-\Lambda_n) - (\pm\rho-\epsilon)a_n}{2}
    \Big\langle \frac1R \Big\rangle_n  , \label{dK4}\\
    a_{n+1} &=& a_n + \Big\langle\frac{da}{d\tau}\Big\rangle_n 
    = a_n+ \frac{\epsilon \beta}{2}  -\frac{\epsilon(a_n)}{2} \Big\langle \frac1R \Big\rangle_n    \label{dA4},\\
    H_{n+1}  &=& H_n+\Big\langle\frac{dH}{d\tau}\Big\rangle_n 
    = H_n+\frac{(\pm\rho -\epsilon\beta)}2 \Big\langle \frac1R \Big\rangle_n 
    \nonumber \\
    && \qquad\qquad\qquad\qquad\qquad\qquad+ \frac{\pm\rho (\lambda-\Lambda_n) - (\pm\rho-\epsilon)a_n}{2} 
    \Big\langle \frac1{R^2} \Big\rangle_n  , \label{dH4}
\end{eqnarray}
where, e.g, $\langle \frac1R \rangle_n = \int_n^{n+1}\frac1{R(\tau)}d\tau$ with $R(\tau)$ given by \eqref{rtau} with $r_0$ and $R_0$ calculated using $H_n$ and $\Lambda_n$. These averages can be found explicitly as
\begin{eqnarray}
    \Big\langle \frac1{R} \Big\rangle_n &=& \frac{\pi}{\sqrt{(\pi r_0)^2 - R_0^2}} = \frac{\pi}{\sqrt {\Lambda_n}} , \\
    \Big\langle \frac1{R^2} \Big\rangle_n &=&\frac{\pi^3r_0}{\sqrt{\big((\pi r_0)^2 - R_0^2\big)^3}}= \frac{\pi{H_n}}{2\sqrt{\Lambda_n^3}}.
\end{eqnarray}
Inserting these expressions in \eqref{dK4}-\eqref{dH4}  yields \eqref{dK5}-\eqref{dH5}.

\end{document}